\documentclass[aps,prb,twocolumn,showpacs,superscriptaddress]{revtex4-2}
\usepackage{amsfonts}
\usepackage{amsmath}
\usepackage{amssymb}
\usepackage{graphicx}
\usepackage{float}
\usepackage{multirow}
\usepackage{ulem}
\usepackage{gensymb}
\usepackage{dcolumn}
\usepackage[dvipsnames]{xcolor}

\def\iu{\mathrm{i}}

\usepackage[bookmarks=false]{hyperref} 
\hypersetup{colorlinks=true,
            linkcolor=blue,
            citecolor=blue,
            urlcolor=orange}
\begin{document}
\title{Charge and spin correlations in insulating and incoherent metal states \texorpdfstring{\\of twisted bilayer graphene}{}}
\author{A. A. Katanin}
\affiliation{Center for Photonics and 2D Materials, Moscow Institute of Physics and Technology, Institutsky lane 9, Dolgoprudny, 141700, Moscow region, Russia}
\affiliation{M. N. Mikheev Institute of Metal Physics of Ural Branch of Russian Academy of Sciences, S. Kovalevskaya Street 18, 620990 Yekaterinburg, Russia}

\begin{abstract}
We study electronic, charge, and magnetic properties of twisted bilayer graphene with fillings $2\leq n\leq 6$ per moire unit cell within the recently introduced  formulation of extended dynamical mean-field theory (E-DMFT) for two-sublattice systems. We use previously obtained hopping parameters between the states, described by Wannier functions, centered at the lattice spots of AB and BA stacking, and the long-range Coulomb interaction, obtained within cRPA analysis. We show that account of spin exchange between AB and BA nearest neighbor spots is crucial to introduce charge and spin correlations between these spots. Account of this exchange yields preferable concentration of electrons in the same valley, with the tendency of parallel spin alignment of electrons in AB and BA spots, in agreement with earlier results of the strong coupling analysis, suggested SU(2)$\times$SU(2) emergent spin-valley symmetry. The local spectral functions show almost gapped state at the fillings $n=2,4,6$, and incoherent metal state for the other fillings. We find that in both cases the local states of electrons have rather long lifetime. At the same time, the non-local charge- and spin susceptibilities, obtained within the ladder approximation, are peaked at incommensurate wave vectors, which implies that the above discussed ordering tendencies are characterized by an incommensurate pattern.
\end{abstract}
\maketitle
Twisted bilayer graphene (TBG), which represents two sheets of graphene, rotated by a small angle with respect to each other, was synthesized experimentally in 2018 \cite{Expt1,Expt2}. This material possesses fascinating properties, showing insulating behavior at the electron fillings of narrow bands $n=2,6$ per moire unit cell  (corresponding to the carrier concentration $\pm 2$ per unit cell) \cite{Expt1,Expt2}, as well as the half filling \cite{Expt3,ExptSTM,ExpTransp,ExptCascade,Cascade}, and superconductivity in the vicinity of the fillings $n=2,6$ \cite{Expt2}. The electron spectral functions measurements by scanning tunneling microscopy (STM) at various fillings \cite{ExptSTM,ExptCascade,Expt5} show clear signature of interaction effects. Explaining features of the spectral functions, observed in these experiments, represents an important theoretical problem.

The peculiarities of the band structure of TBG were discussed long time before its experimental realization \cite{ES1,ES2,ES3,ES4,ES5,ES6,ES7} and studied in detail soon after it
\cite{ES8,ES9,ES10,ES11,ES12,ES13,ES14,ES15}. At small twist angles almost flat electronic bands are formed, with the bandwidth of the order of 10 meV, which depends however on the twist angle \cite{ES9,Int1}. The corresponding Wannier functions
are formed by the electronic states centered at the spots of the lattice with AB and BA stacking \cite{ES8,ES9,ES10,ES11,ES12}, the hopping parameters between these spots were obtained by means of Wannier projection \cite{ES10,ES11}.

Since the electronic dispersion of TBG possesses Dirac points, the screened Coulomb interaction remains long range (see, e.g., Refs. \cite{Screening,IntcRPA1}). The matrix elements of this interaction between Wannier states were determined in Refs. \cite{ES10,SU41}. The intra-spot Coulomb repulsion in the presence of substrate with the dielectric permittivity $\epsilon=5$ is estimated as $V_0\simeq 38$~meV for the twist angle $\theta=1.05\degree$ \cite{ES10}. This value of $V_0$ is larger than the bandwidth of narrow bands, which implies a possibility of interaction-induced Mott metal insulator transition; similar result for $V_0$ was obtained for
$\theta=1.08\degree$ \cite{SU41}. The screening of the interaction by the other bands was investigated within the cRPA approach \cite{IntcRPA1,IntcRPA2,IntcRPA3,Int2} and reduces the above mentioned intra-spot repulsion to $V_0\simeq 15$~meV for the  permittivity of the substrate $\epsilon=5$ and $\theta=1.05\degree$ \cite{IntcRPA2}; close value was obtained for $\theta=1.08\degree$, $\epsilon=4$ in Ref. \cite{IntcRPA3}. Therefore, for realistic parameters screened interaction remains larger than the bandwidth. 

The effect of these interactions on the phase diagram was studied within the weak coupling approaches, such as the random phase approximation \cite{MF6,MF7} and renormalization group \cite{RG1,RG2,RG3,RG4}, strong coupling approaches \cite{SU41,SU42}, as well as the approaches not formally restricted by the interaction strength, in particular mean-field approach \cite{MF1,MF2,MF3,MF4,MF5}, Monte Carlo \cite{MC}, exact diagonalization \cite{ED}, and dynamical mean field theory (DMFT) \cite{MFDMFT}. These approaches yielded variety of phases, including ferromagnetism (see also Refs. \cite{FM1,FM2,FM3}), spin density waves, valence bond order (see also Refs. \cite{VBS1,VBS2,VBS3}), etc. Special emphasis was paid to the emergent SU(2)$\times$SU(2) spin and valley symmetry in the strong coupling limit\cite{SU41,SU42,MF1}, which yields formation of mixed valley-spin ordered states with ferromagnetic alignment of the spins of the same valley.

Although the possibility of Mott transition in TBG was emphasized right after its syntesis in Refs. \cite{Expt1,Expt2,Expt3,ExptSTM,ExpTransp,ExptCascade,Cascade,ES8,Expt5}, only several theoretical approaches are able to treat this possibility. In view of sufficiently strong Coulomb interactions in TBG, discussed above, the dynamical mean-field theory \cite{DMFT_rev}, including the extended DMFT (E-DMFT) approach \cite{EDMFT_Si,EDMFT,EDMFT,EDMFT1}, as well as their non-local diagrammatic extensions \cite{OurRev,MyEDMFT,DB,AbInitioDGA,MyEDMFTfRG}, are suitable tools for treatment of both, local and non-local interactions in this system. Previously, mainly only on site Coulomb interaction was considered in DMFT studies of TBG \cite{Expt5,MFDMFT}. 

Recently, the formulation of the E-DMFT approach for multi-sublattice systems was used \cite{MyEDMFT} to study charge and spin correlations in graphene. This approach treats explicitly both, local and the non-local interaction inside the unit cell. The remaining part of the non-local interaction was considered by an effective retarded intra unit cell interaction of E-DMFT. The non-local charge and spin susceptibilities can be further considered within the ladder non-local diagrammatic extensions of E-DMFT approach \cite{OurRev,MyEDMFT,DB,AbInitioDGA,MyEDMFTfRG}. The results of the above described E-DMFT method for graphene \cite{MyEDMFT} showed good agreement with the results of functional group approach \cite{OurFlakes}, as well as previous results of quantum Monte-Carlo studies. 

In the present paper we apply the above described approach to investigate the electronic properties, and study the effect of charge and spin correlations in TBG. In contrast to the earlier studies of Refs. \cite{Expt5,MFDMFT}, we consider the effect of the long range Coulomb interaction obtained within the cRPA analysis \cite{IntcRPA2} for tight-binding model of electrons with Wannier functions centered at AB, BA spots.
We also account for the magnetic exchange interaction between AB and BA spots, discussed in Refs. \cite{ES10,SU41}. We show that the effect of the latter interaction is crucial to resolve between different types of correlations and find dominating charge susceptibility, which is odd in valley index, but even in sublattice (AB and BA) indexes, such that electrons prefer to concentrate in the same valley but fill almost equally AB and BA spots. The dominating spin susceptibility is even in sublattice index, showing preferable ferromagnetic ordering of nearest neighbor sites. We obtain local spectral functions at various fillings, and consider other local and non-local properties. 

\begin{figure}[t]
		\center{
				\vspace{0.3cm}
		\includegraphics[width=0.8\linewidth]{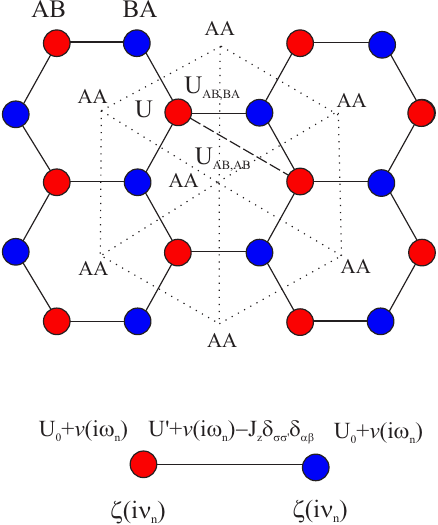}}
		\caption{
		{(Color online).} Upper part: Fragment of the hexagonal lattice of AB and BA spots with the on-site ($U_0$), nearest neighbor ($U_{{\rm AB},{\rm BA}}$, solid lines), next nearest neighbor ($U_{{\rm AB},{\rm AB}}$)  interactions (dashed line), longer distance interactions are considered within the model (\ref{GNFHamiltonian}), but not shown. Dotted lines show the lattice of AA spots. Lower part: the impurity model formed by ${\rm AB}$ and ${\rm BA}$ spots, including the bath Green functions $\zeta(\iu \nu_n)$, acting at each spot, the on-site interaction $U_0$, the interspot interaction $U'$ (shown by solid line), magnetic exchange $J_z$, and the dynamic interaction $v(i\omega_n)$.}
\label{SystemsPic}
\end{figure}

\vspace{0.2cm}
{\it Model and method}. To model properties of TBG, we consider the tight-binding model of electrons, described by the Wannier functions, centered at AB and BA spots on a hexagonal lattice, with the hopping between different spots and long-range interaction (see Fig. \ref{SystemsPic}). The corresponding Hamiltonian can be written as
\begin{align}
 \mathcal{H}&=-\sum_{im\alpha,jm'\beta,\sigma}t_{im\alpha,jm'\alpha}\left({\hat d}^{\dagger}_{im\alpha\sigma}{\hat d}_{jm'\alpha\sigma}+H.c.\right)\label{GNFHamiltonian}\\&+\dfrac{1}{2}\sum_{im,jm',\alpha\beta}U^{\sigma\sigma'}_{im\alpha,jm'\beta}\left({\hat n}_{im\alpha\sigma}-\frac{1}{2}\right)\left({\hat n}_{jm'\sigma\beta}-\frac{1}{2}\right).\notag
\end{align}
Here, ${\hat d}^{\dagger}_{im\alpha\sigma}$ $({\hat d}_{im\alpha\sigma})$ is a creation (annihilation) operator of an electron at the unit cell $i$ of the hexagonal lattice, $m={\rm AB,{\rm BA}}$ is the spot index, $\alpha=1,2$ and $\sigma=\uparrow,\downarrow$ are the valley  and spin indexes, ${\hat n}_{jm\alpha\sigma}={\hat d}^{\dagger}_{jm\alpha\sigma}{\hat d}_{jm\alpha\sigma}$. We consider twist angle $\theta=1.05\degree$ and take the hopping parameters $t_{im\alpha,jm'\alpha}$ from the Wannier projection of the continuum model in Ref. \cite{ES10}.
 The last term in Eq.~(\ref{GNFHamiltonian}) describes the electron-electron interaction with the potential $U^{\sigma\sigma'}_{im\alpha,jm'\beta}$  
 that includes both the on-site and non-local contributions. 
 
\begin{table*}[t!]
\begin{tabular}{||l||ldddd||ldd||ddd||}
\hline\hline
 & \multicolumn{5}{c||}{$\chi^c_{\rm loc}$}& \multicolumn{3}{c||}{$\chi^s_{\rm loc}$} & \multicolumn{3}{c||}{$\langle n_{{\rm AB}_{1}\uparrow}n_{m\alpha\uparrow}\rangle$}\\ \hline
$m_\alpha$
&& \multicolumn{1}{l}{${\rm AB}_{1}$} & \multicolumn{1}{l}{${\rm BA}_{1}$} & \multicolumn{1}{l}{${\rm AB}_{2}$} & 
\multicolumn{1}{l||}{${\rm BA}_{2}$} 
&\,\,& \multicolumn{1}{l}{${\rm AB}_{1}$} & \multicolumn{1}{l||}{${\rm BA}_{1}$} 
& \multicolumn{1}{r}{${\rm BA}_{1}$} & \multicolumn{1}{r}{${\rm AB}_{2}$} & \multicolumn{1}{c||}{${\rm BA}_{2}$}
\\ \hline\hline
 $n=2$, $J_{z}=0$ &
  \,\,\, &$1.29$ & $-0.04$ & $-1.11$& $-0.04$
 && $2.41$ & $0$ &
 0.06 &0.01 &0.06\\ 
$n=2$, $J_{z}$ 
&
 \, & $1.25$ & $1.12$ & $-1.23$ & $-1.12$ 
&&$2.48$ & $2.24$
&0.23&0.00&0.01 \\ 
$n=3$, $J_{z}$ 
& \, & $1.75$ & $1.08$ & $-1.13$ & $-1.04$
 && $2.88$ & $2.12$ 
 &0.30&0.08&0.09\,\,\,\,\\ 
$n=4$, $J_{z}$ 
& 
 \, & $1.67$ & $1.49$ & $-1.64$ & $-1.50$ 
&& $3.31$ & $2.99$ 
&0.47&0.17&0.17\\ 
\hline \hline
\end{tabular}
\caption{Charge (spin) static local susceptibility $\chi_{\rm loc}^{c(s) {\rm AB}_1,m\alpha}$ and double occupations for integer fillings. The inter valley local spin susceptibility ($\alpha=2$) is negligibly small, $\langle n_{{\rm AB}_{1}\uparrow}n_{m_{\alpha}\downarrow}\rangle\approx \langle n_{{\rm AB}_{1}\uparrow}n_{m_{2}\uparrow}\rangle$. }
    \label{Tabchiloc}
\end{table*}
 
 For the following discussion we split the interaction $U^{\sigma\sigma'}_{im\alpha,jm'\beta}=U_{im,jm'}+\Delta U_{im,jm'}\delta_{\alpha\beta}\delta_{\sigma\sigma'}$. The valley independent  and and spin isotropic part was obtained within cRPA analysis \cite{IntcRPA2}. It can be parameterized by 
 $U_{im,jm'}=U_0/(1+U_0/W(r_{im,jm'}))$, where $U_0=15$~meV is the on-site interaction, $W(r)=e^2/(\epsilon r)$ is the bare Coulomb repulsion, and ${\bf r}_{im,jm'}$ is the radius-vector connecting corresponding lattice sites $i,m$ and $j,m'$. According to Refs. \cite{ES10,SU41} we also include the intravalley nearest neighbor ferromagnetic exchange $\Delta U_{i{\rm AB},j{\rm BA}}=-J_z$ where we choose $J_z=3.75$~meV \cite{NoteJz}. Because of the limitations of the used impurity solver, we neglect exchange interaction at the distances longer than the nearest neighbors distance, which is justified by sufficiently fast decay of this interaction with the distance \cite{SU41,ES10}.
 We also consider only intravalley longitudinal $z$-component $J_z$ of the spin interaction, since only this part of the interaction can be reduced to the density-density form, allowed by the impurity solver. This approximation corresponds to breaking $SU(2)$ spin symmetry and the emergent (in the strong coupling limit) $O(2)\times Z_2$ valley symmetry (see Ref. \cite{SU41}) to the $Z_{2,\rm spin}\times Z_{2,\rm valley}$ one. We note that neglect of the transverse part of the exchange (e.g. Hund) interaction is a rather common approximation in DMFT studies of multi-band systems (see, e.g., Ref. \cite{DFTDMFT}), and it is known to yield an overestimate of phase transition temperatures, while capturing the main physical properties of the system. We therefore expect that the main ordering tendencies are captured by included interactions.

%

Following Ref. \cite{MyEDMFT}, we Fourier transform the isotropic part of the interaction $V_{mm'}({\bf q})=\sum_{j} U_{im,jm'}e^{\iu {\bf q} {\bf r}_{im,jm'}}$, and introduce averaged intra- ($U_{{\rm AB},{\rm AB}}=U_0$) and intersublattice ($U_{{\rm AB},{\rm BA}}=U'$) interaction over momentum, $U_{mm'}=\sum_{\bf q} V_{mm'}({\bf q})$. The remaining non-local isotropic interaction ${\widetilde V}_{mm'}({\bf q})=V_{mm'}({\bf q})-U_{mm'}$ is considered within the E-DMFT approach \cite{EDMFT_Si,EDMFT,EDMFT1} by introducing the self-consistently determined effective dynamic interaction $v(\iu\omega_n)$ in the impurity model and accounting for the difference ${\widetilde V}_{mm'}({\bf q})-v(\iu\omega_n)$ in the ladder summation for susceptibilities (see details in Refs. \cite{MyEDMFT,SM}). The anisotropic part of the interaction is introduced in the impurity model, which reads 
\begin{align}
S&_{\rm DMFT}=-\sum_{m\alpha,\iu \nu_n}  \zeta^{-1}(\iu \nu_n) d^{\dagger}_{im\alpha\sigma}(\iu \nu_n)d_{im\alpha\sigma}(\iu \nu_n)\notag\\
&+\frac{1}{2}\sum_{m\alpha\sigma,m'\beta\sigma',\iu\omega_n}U^{\sigma\sigma'}_{m\alpha, m'\beta}  {n}_{im\alpha\sigma}(\iu \omega_n)n_{im'\alpha\sigma'}(-\iu \omega_n)\notag\\
&+\frac{1}{2}\sum_{\iu\omega_n}v(\iu\omega_n)  n_{i}(\iu \omega_n) n_{i}(-\iu \omega_n),
\label{Simp}
\end{align}
where ${ d}^{\dagger}_{im\alpha\sigma}$ $({d}_{im\alpha\sigma})$ are Grassmann variables, $i$ refers to the impurity site,  $U_{m\alpha,m\beta}^{\sigma\sigma}=U_0$ ($\alpha\ne \beta$), $U_{m\alpha,m\beta}^{\sigma,-\sigma}=U_0$, $U_{{\rm AB}_\alpha,{\rm BA}_\beta}^{\sigma\sigma'}=U_{{\rm BA}_\alpha,{\rm AB}_\beta}^{\sigma\sigma'}=U'-J_z\delta_{\sigma\sigma'}\delta_{\alpha,\beta}$ (when $m$ is specified explicitly we denote the combination $m\alpha$ as $m_\alpha$). The bath Green function $\zeta(\iu \nu_n)=\{[G^{\rm loc}(\iu \nu_n)]^{-1}+\Sigma(\iu \nu_n)\}^{-1}$ 
is determined self-consistently, $n_{im\alpha\sigma}(\iu \omega_n)=\sum_{\nu_n} { d}^{\dagger}_{im\alpha\sigma}(\iu \nu_n) {d}_{im\alpha\sigma} (\iu \nu_n+\omega_n)$, $n_{i}(\iu \omega_n)=\sum_{m\alpha\sigma}n_{im\alpha\sigma}(i \omega_n)$. 


For the solution of the impurity problem we apply the continous-time quantum Monte-Carlo (CT-QMC) approach, realized in the iQIST package \cite{iQIST}. In view of approximate particle-hole symmetry of the dispersion of Ref. \cite{ES10}, we mainly consider interval of the fillings of electrons $2\leq n\leq 4$ per moire unit cell ($n=4$ corresponds to half filling); we have verified that the results for the fillings $4<n\leq 6$ are close to those obtained by applying particle-hole transformation \cite{NotePH,SM}.  The calculations are performed at $T=11.6$~K, which is approximately 10 times smaller than the band width, and $\epsilon=5$. In view of not too low considered temperature, we perform calculations in the spin-, spot-, and valley-symmetric state.




{\it Results}. Let us first analyse the results for the local charge and spin susceptibility $\chi_{\rm loc}^{c(s) m'\beta,m\alpha}(\omega)=(1/2)\langle\langle \rho^{c(s)}_{im'\beta}|\rho^{c(s)}_{im\alpha}\rangle\rangle_{\omega}$ at $\omega=0$, 
describing long-time charge and spin local correlations,
where $\rho^{c}_{im\alpha}=\sum_{\sigma} n_{im\alpha\sigma}-n/4$, $\rho^{s}_{im\alpha}=\sum_{\sigma}(- 1)^{\sigma}n_{im\alpha\sigma}$. In the Table \ref{Tabchiloc} we present the results for the susceptibilities and double occupations $\langle n_{m'\beta\uparrow}n_{m\alpha\sigma}\rangle$ at integer fillings for $m',\beta={\rm AB}_1$ (the results for the other spot and valley indexes can be obtained by symmetry; here and hereafter if not specified otherwise explicitly, we use energy units of $10$~meV). For $n=2$ we present the results with and without magnetic exchange (for other fillings the comparison looks similarly). One can see that without magnetic exchange charge correlations are mostly present only within the same spot, while spin correlations - within the same spot and valley. Accordingly, the occupations and spin orientations of different spots are independent in this case, and they can also be different in different valleys. With inclusion of magnetic exchange charge correlations are spread to both, spots and valleys and indicate a tendency towards filling of one of the two valleys, according to positive (negative) intra (inter) valley charge correlations. The above mentioned tendency is especially pronounced for $n=2$, when the double occupations of electrons with different valley  or spin index almost vanish. At larger $n$ finite double occupations of different valleys or spin states occur,
although such occupations are still suppressed. While the intra- and inter spot charge susceptibilities within the same valley are almost equal for $n=2$ and $n=4$, reflecting equal preferable occupation of these spots, for $n=3$ stronger imbalance of the intra valley charge susceptibilities is also observed. The spin correlations in the presence of magnetic exchange involve both spots of the same valley, indicating preferable ferromagnetic alignment of the spin states at these spots. Therefore, we observe crucial effect of magnetic exchange on charge and spin correlations, which support valley- and spot states, similar to discussed previously within strong coupling analysis of Refs. \cite{SU41,SU42}. One can also see that the local spin correlations are enhanced on approaching half filling $n=4$.

\begin{figure}[t]
	\center{\includegraphics[width=1.0\linewidth]{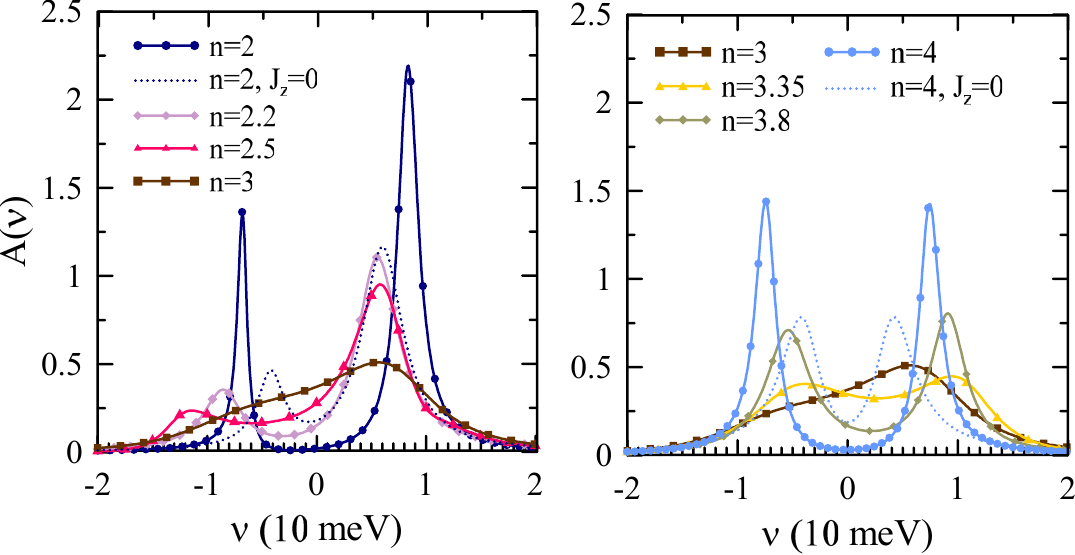}}
			\caption{(Color online) The frequency dependence of the local spectral functions at various fillings. Dotted lines show the results for $J_z=0$ and $n=2$ (left plot) and $n=4$ (right plot).}
	\label{Aw}
\end{figure}

The local spectral functions $A(\nu)=(-1/\pi){\rm Im}G_{\rm loc}(\nu)$, obtained by analytical continuation of E-DMFT local Green's function using Pade approximants, are shown for various fillings in Fig. \ref{Aw}. As we explicitly show in Supplemental Material, the results for the fillings $n=5,6$ are close to those for $n=2,3$, up to the particle-hole transformation. One can see that for fillings $n=2,4,6$ the spectral functions are almost gapped at the Fermi level ($\nu=0$) due to strong electronic correlations, and correspond to the insulating states, which agrees with the experimental data \cite{Expt1,Expt2,Expt3,ExptSTM,ExpTransp,ExptCascade}. In agreement with the discussion above, one can see that switching off $J_z$ yields more metallic spectral functions. The peaks of the spectral functions correspond to broadened levels of impurity problem, which in the absence of the retarded part $v(i\omega_n)$ are located at $\pm (U_0+J_z)/2$ (see Supplemental Material \cite{SM}), and slightly shifted by the retarded interaction. We note that earlier the peaks of the spectral functions, obtained experimentally \cite{vHSExpt}, were associated with the extended (higher-order) van Hove singularities \cite{vHSExpt,vHS}. However, the experimental distance between the peaks $\simeq 57$~meV at $\theta=1.1\degree$ is too large in comparison to the bandwidth $\sim 10$~meV, estimated from the ab initio approaches,  which required to readjust bandwidth in Ref. \cite{vHSExpt}. For $n\ne 2,4,6$ we obtain the non-zero density of states at the Fermi level; the maximum of the spectral function at $n=3,5$ corresponds to the broadened atomic level $\pm(U'-J_z)/2$ (see Supplemental Material \cite{SM}). As we discuss in Supplemental Material \cite{SM}, even in this case the obtained frequency dependence of the electronic self-energy has a non-quasiparticle form. This state can be therefore characterized as an incoherent metallic state.

\begin{figure}[t]
	\center{\includegraphics[width=0.8\linewidth]{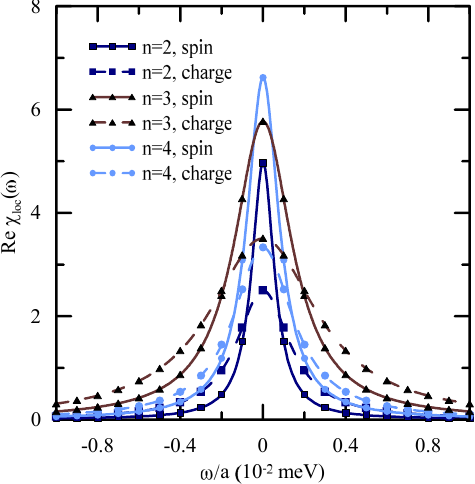}}
			\caption{(Color online) The frequency dependence of the real part of local charge- (dashed lines) and spin (solid lines)  susceptibilities $\chi_{\rm loc}^{c(s) m\alpha,m\alpha}(\omega)$ at the real frequency axis and  integer fillings. The rescaling factor $a=1$ for $n=2,4$, and $a=10$ for $n=3$ is introduced for visibility.}
	\label{chiw}
\end{figure}

To study further the degree of electron localization at integer fillings, we present in Fig. \ref{chiw} the frequency dependence of the real parts of the diagonal local spin- and charge- susceptibilities $\chi_{\rm loc}^{c(s) m\alpha,m\alpha}(\omega)$ at the real frequency axis, obtained by Pade analytical continuation of E-DMFT results. The narrow peaks of the real part show charge and spin localization, with the local states lifetime $\tau$, given by the inverse width of the peak of the real part, cf. Refs. \cite{OurAlpha,OurGamma,Toschi1,Toschi2,Toschi3}). For $n=2$ and $n=4$ we obtain $\tau\sim 5$~ns, while for $n=3$ we have $\tau\sim 0.5$~ns; in all cases we find $\tau\gg h/(k_B T)$ ($h$ and $k_B$ are Planck and Boltzmann constants, respectively). Importantly, even in the incoherent metallic state at $n=3$ this is a rather long local state lifetime, in comparison to the typical lifetimes of local magnetic moments obtained in such strongly correlated substances as pnictides ($\tau\sim 10$~fs, Ref. \cite{Toschi1}), and even $\alpha$-iron ($\tau\sim h/(k_B T)\sim 5$~ps at the considered temperature \cite{OurAlpha,OurGamma}).
Even longer lifetimes of local states of TBG are expected at lower temperatures.

\begin{figure}[t]
	\center{
\includegraphics[width=1.0\linewidth]{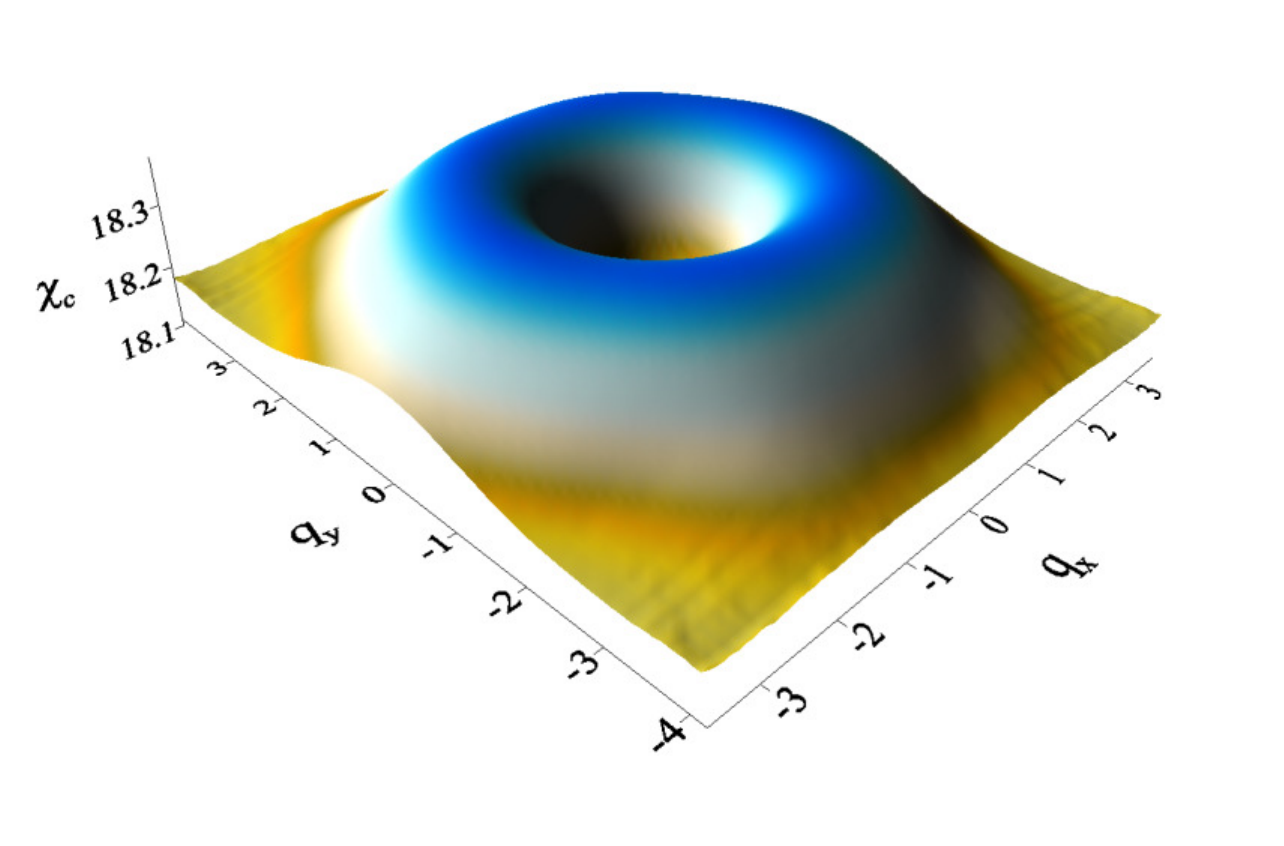}}
			\caption{(Color online) The momentum dependence of the staggered in valley indexes charge susceptibility $\chi_{\bf q}^{c,{\rm st}}$ at $n=2$. The momentum dependence of the spin susceptibility is quite similar, see Supplemental Material \cite{SM}.}
	\label{chic}
\end{figure}

To study the non-local spin- and charge correlations we calculate the non-local static charge (spin) susceptibility 
$\mathcal{\chi}_{\bf q}^{c(s),m\alpha,n\beta}=(1/2)\langle\langle \rho^{c(s)}_{{\mathbf q},m\alpha}|\rho^{c(s)}_{-{\mathbf q},n\beta}\rangle\rangle_{\omega=0}$,
where $\rho^{c(s)}_{{\mathbf q},m\alpha}$ is the Fourier transform of $\rho^{c(s)}_{im\alpha}$. These susceptibilities are evaluated in the ladder approximation via the numerical solution of the Bethe-Salpeter equation, using local vertices, obtained within the E-DMFT approach, see Ref. \cite{MyEDMFT} for the details (cf. also Refs. \cite{DB,AbInitioDGA,OurRev,My_BS}). For calculation of the local vertices within the CT-QMC method we use 40-60 fermionic frequencies (both positive and negative). In Fig. \ref{chic} we show the resulting momentum dependence of the staggered with respect to valleys charge susceptibility $\chi_{\bf q}^{c,{\rm st}}=\sum_{mn\alpha\beta}(-1)^{\alpha+\beta} \mathcal{\chi}_{\bf q}^{c,m\alpha,n\beta}$ for $n=2$. The considered susceptibility is dominant among other uniform/staggered charge susceptibilities in view of the analysis of the local counterpart, presented above; the wave vectors are shown in the units of $L_{AA}^{-1}$ where $L_{AA}=a_{\rm tr}/(2\sin(\theta/2))$ is the supercell lattice constant of $AA$ spots, and $a_{\rm tr}=2.46\text{\AA}$ is the lattice constant of the sites of one of the graphene's sublattices. We find that the most preferable ordering tendency corresponds to an incommensurate pattern with a continuous set of the wave vectors, forming almost a circle in momentum space with the radius close to $2L_{AA}^{-1}$, which implies periodicity in the real space with the period $\sim 3L_{AA}$. Surprisingly, we find almost the same pattern in the spin susceptibility $\chi_{\bf q}^{s}=\sum_{mn\alpha\beta} \mathcal{\chi}_{\bf q}^{s,m\alpha,n\beta}$, with slightly different maximal value, see Ref. \cite{SM}. Therefore, the space distribution of both, charge and spin correlations in TBG at $n=2$ is expected to be the same, and characterized by the obtained set of the incommensurate wave vectors, which is another consequence of the emergent charge-spin symmetry. Similar results are obtained for the other fillings, the examples of momentum dependencies of susceptibilities for the integer fillings $n>2$ are presented in the Supplemental Material \cite{SM}. With approaching half filling of the moire unit cell ($n=4$) the charge and spin susceptibilities increase, while the wave vector of incommensurate correlations decreases, such that the charge and spin correlations become more commensurate. 

{\it Conclusion}. In this paper, we have applied recently developed formulation of E-DMFT approach \cite{MyEDMFT} to consider the electronic states, local and non-local charge and spin correlations of TBG in the filling range $2\leq n\leq 6$. We have used previously obtained Wannier projected dispersion of electrons, moving on a lattice of the spots with AB and BA stacking  and cRPA screened interaction, which includes both, Coulomb repulsion and magnetic exchange. In the presence of magnetic exchange between nearest-neighbor AB and BA spots we find the tendency of electrons to occupy the same valley, and fill almost equally the nearest neighbor spots. The magnetic exchange favors also ferromagnetic alignment of spins of nearest neighbor spots. The effect of this exchange is crucial for obtaining the above discussed state; without the magnetic exchange the correlations between different spots become negligibly small. 

The obtained state in the presence of magnetic exchange for $n=2(6)$, when the double occupation of electrons (holes) is present only within the same valley, is similar to that earlier discussed in the strong coupling analysis \cite{SU41,SU42} within emergent SU(2)$\times$SU(2) spin valley symmetry scenario. With approaching half filling the double occupation of different valleys occurs, although it is suppressed by correlations. At lower temperatures this can yield cascade of phase transitions, which is similar to that discussed recently in Ref. \cite{Cascade}, with the difference that we expect a tendency to the equal occupation of the spots within the same valley instead of filling equally different spin projections. Yet, at the considered temperature we find finite local and non-local charge and spin susceptibilities, such that the spontaneous symmetry breaking does not occur.

The local spectral functions, obtained within E-DMFT analysis, show gapped state for the fillings $n=2,4,6$ and incoherent metal state for the other fillings. The obtained spectral functions qualitatively agree with the STM study of Ref. \cite{Expt5}. The spectral functions in the vicinity of the fillings $n=4,6$ also qualitatively agree with the studies \cite{ExptSTM,ExptCascade}. The disagreement at some other fillings, as well as mutual disagreement between some features of the above mentioned STM studies require further clarification, but can be at least partly explained by the tip-induced band bending (see discussion in Ref. \cite{ExptSTM}). At all fillings we find that local magnetic and charge states have rather large lifetime of the order of few nanoseconds at the considered temperature $T=11.6K$; even longer life times are expected at lower temperatures. 

Based on the solution of E-DMFT problem, the local vertices were calculated and the non-local charge and spin susceptibilities were evaluated via the solution of the Bethe-Salpeter equation. In both, staggered in valleys charge channel and in the spin channel we find incommensurate pattern of preferable ordering tendencies with the wavevector $\sim 2L_{\rm AA}^{-1}$, and the corresponding real space periodicity at distances $\sim 3L_{\rm AA}$ where $L_{\rm AA}$ is the supercell lattice constant of AA spots. 

In the considered approach we have accounted
only intravalley longitudinal $z$-component $J_z$ of the spin interaction, since only this part can be reduced to a density-density form, allowed by the used impurity solver. 
 We expect that this does not change qualitatively the obtained results, since main effect on the ordering tendencies is captured by the included interactions. We have also included the magnetic exchange only between the nearest neighbor AB and BA spots in view of its fast decay with distance; considering longer range magnetic exchange requires treatment of the dynamic spin interaction in the impurity problem. Using solvers, which account for the transverse spin and/or isospin valley part of the exchange interaction for the considered 4-band model is more challenging problem, which can be considered in future studies.

The developed method can be further used to study superconductivity of TBG near integer fillings. Another interesting topic is studying dynamic collective excitations, such as magnons, plasmons, etc. in twisted bilayer graphene, as well as for studying other related systems.

In view of strong correlations in TBG, an interesting task for future studies is also considering the non-local corrections to the self-energy, which will allow to study an effect of the renormalization of the Fermi velocity, damping of electronic quasiparticles due to non-local correlations, and can be performed within one of the diagrammatic extensions of E-DMFT approach, cf. Refs. \cite{DB,AbInitioDGA,OurRev,MyEDMFTfRG}.

{\it Acknowledgements}. The author acknowledges the financial support from the Ministry of Science and Higher Education of the Russian Federation (Agreement No. 075-15-2021-606 and theme "Quant" AAAA-A18-118020190095-4). The work is also partly supported by RFBR grant 20-02-00252 A. The calculations were performed on the cluster of the Laboratory of material computer design of MIPT and the Uran supercomputer at the IMM UB RAS.

\clearpage
\begin{widetext}
\appendix
\renewcommand\theequation{A\arabic{equation}}
\renewcommand\thefigure{S\arabic{figure}}
\renewcommand\thetable{S\arabic{table}}
\setcounter{equation}{0}
\setcounter{figure}{0}
\setcounter{table}{0}
\section*{Supplemental Material 
\texorpdfstring{\\ \lowercase{to the paper }``C\lowercase{harge and spin correlations in insulating and incoherent metal states \\of twisted bilayer graphene}"}{}}
\vspace{-0.4cm}
\centerline{A. A. Katanin}
\vspace{0.5cm} 

\subsection{Electronic spectral functions and dynamic susceptibility for $n=5,6$}

In Fig. \ref{Achiw56} we present the electronic spectral functions and dynamic local charge and spin susceptibilities for $n=5,6$. As it is mentioned in the main text, they are close to the corresponding quantities for $n=2,3$, up to the particle-hole transformation.

\begin{figure}[h!]
	\center{\includegraphics[width=0.45\linewidth]{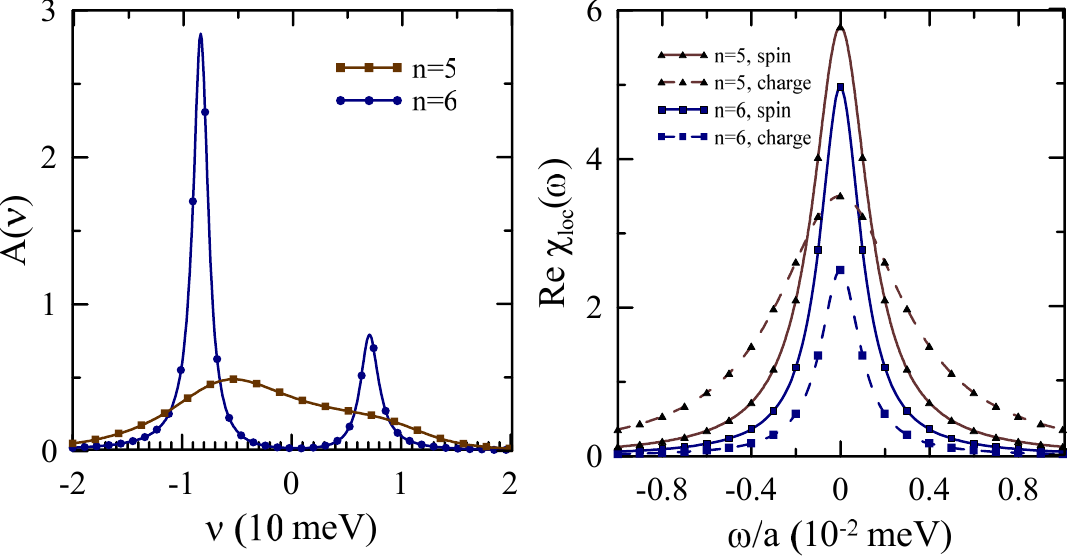}}
			\caption{The frequency dependence of spectral functions (left) and dynamic local spin and charge susceptibilities for $n=5,6$. The factor $a=1$ for $n=6$ and $a=10$ for n=5 is introduced for visibility}
	\label{Achiw56}
\end{figure}

\subsection{Electronic self-energy on the imaginary frequency axis}

\begin{figure}[h!]
	\center{\includegraphics[width=0.45\linewidth]{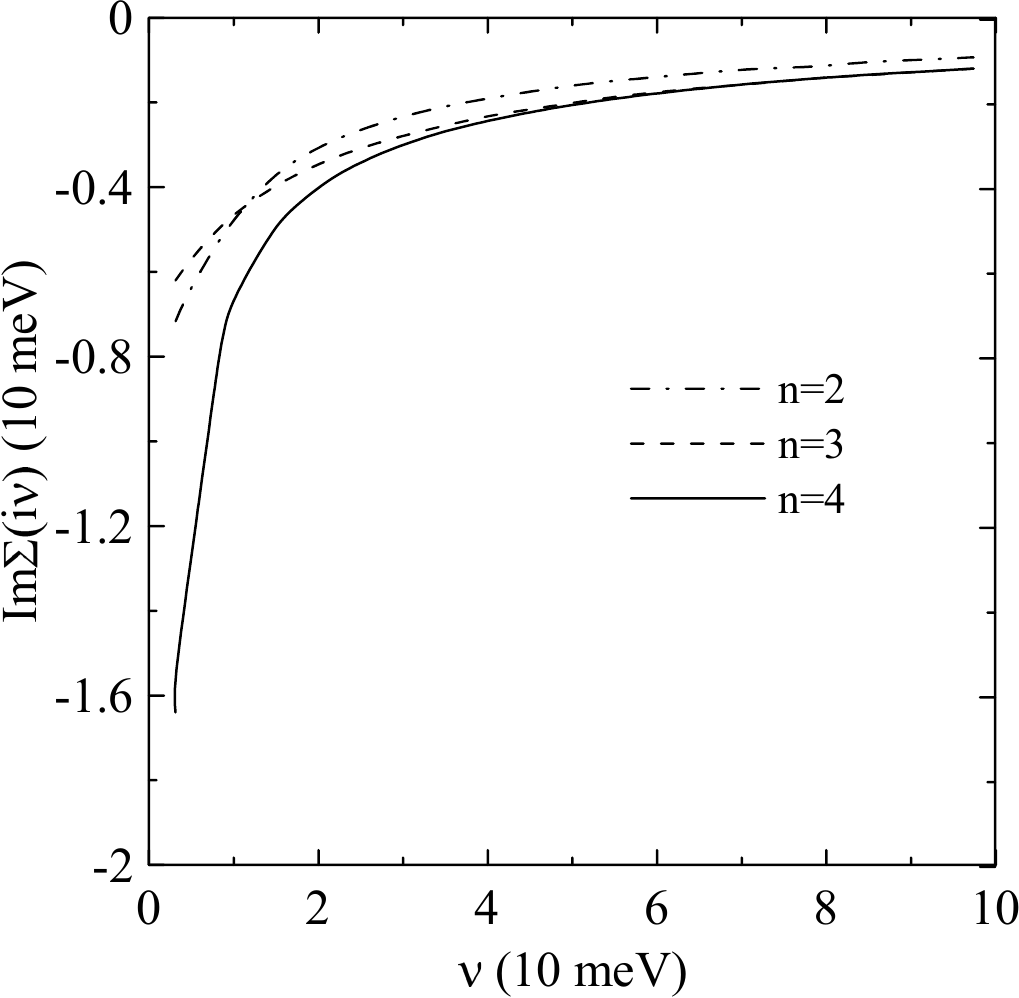}}
			\caption{The frequency dependence of the imaginary part of the self-energy at the imaginary frequency axis for $n=4$ (solid line), $n=3$ (dashed line) and $n=2$ (dot-dashed line).}
	\label{Sigmaw}
\end{figure}

In Fig. \ref{Sigmaw} we present frequency dependence of the imaginary part of the self
energy at the imaginary frequency axis for integer fillings. One can see that for all considered
fillings the derivative $\partial\operatorname{Im}\Sigma(i\nu)/\partial\nu>0,$
which corresponds to the non-quasiparticle states. The damping
$|\operatorname{Im}\Sigma(\nu\rightarrow0)|$ is maximal for $n=4$, intermediate
for $n=2$, and smallest for $n=3,$ being however in all considered cases larger
than the temperature (which is $1$ meV).

\subsection{The energy levels of the atomic problem}

We classify all possible states of the interaction part of the impurity
problem, Eq. (2) of the main text, $|s_{{\rm AB}_{1}},s_{{\rm AB}_{2}},s_{{\rm BA}_{1}%
},s_{{\rm BA}_{2}}\rangle,$ where $s_{m\alpha}=0,\uparrow,\downarrow,\uparrow
\downarrow$ are the corresponding states of the spot $m$ and valley $\alpha,$
by the respective particle number and atomic energy. Neglecting the dynamic
interaction $v(i\omega_{n}),$ which can not be treated analytically, we find the lowest energy states, presented in the Table \ref{En}.%

 \begin{table*}[h!]
\begin{tabular}
[c]{||l|l|c|l||}\hline\hline
$n$ & states (up to the permutations of valleys and spots) & Number of states &
Energy $E_n$\\\hline\hline
0 & $|0,0;0,0\rangle$ & 1 & $0$\\\hline
$1$ & $|\uparrow,0;0,0\rangle,|\downarrow,0;0,0\rangle$ & 8 & $-\mu$\\\hline
$2$ & $|\uparrow,\uparrow;0,0\rangle,|\downarrow,\downarrow;0,0\rangle$ & 4 &
$U^{\prime}-J_{z}-2\mu$\\\hline
$3$ & $|\uparrow\downarrow,\uparrow;0,0\rangle,|\uparrow\downarrow
,\downarrow;0,0\rangle,|\uparrow,\uparrow;\uparrow,0\rangle,|\uparrow
,\uparrow;\downarrow,0\rangle$ & 24 & $U+2U^{\prime}-J_{z}-3\mu$\\\hline
$4$ & $|\uparrow\downarrow,\uparrow\downarrow;0,0\rangle,|\uparrow
,\uparrow;\uparrow,\uparrow\rangle,|\uparrow,\uparrow;\downarrow
,\downarrow\rangle,|\downarrow,\downarrow;\downarrow,\downarrow\rangle$ & 6 &
$2U+4U^{\prime}-2J_{z}-4\mu$\\\hline
$5$ & $|\uparrow\downarrow,\uparrow\downarrow;\uparrow,0\rangle,|\uparrow
\downarrow,\uparrow\downarrow;\downarrow,0\rangle,|\uparrow\downarrow
,\uparrow;\downarrow,\downarrow\rangle,|\uparrow\downarrow,\uparrow
;\uparrow,\uparrow\rangle,|\uparrow\downarrow,\downarrow;\uparrow
,\uparrow\rangle,|\uparrow\downarrow,\downarrow;\downarrow,\downarrow\rangle$
& 24 & $4U+6U^{\prime}-2J_{z}-5\mu$\\\hline
$6$ & $|\uparrow\downarrow,\uparrow\downarrow;\uparrow,\uparrow\rangle
,|\uparrow\downarrow,\uparrow\downarrow;\downarrow,\downarrow\rangle$ & 4 &
$6U+9U^{\prime}-3J_{z}-6\mu$\\\hline
$7$ & $|\uparrow\downarrow,\uparrow\downarrow;\uparrow\downarrow
,\uparrow\rangle,|\uparrow\downarrow,\uparrow\downarrow;\uparrow
\downarrow,\downarrow\rangle$ & 8 & $9U+12U^{\prime}-3J_{z}-7\mu$\\\hline
$8$ & $|\uparrow\downarrow,\uparrow\downarrow;\uparrow\downarrow
,\uparrow\downarrow\rangle$ & 1 & $12U+16U^{\prime}-4J_{z}-8\mu$\\\hline\hline
\end{tabular}
\caption{Lowest energy states for various fillings $n$ and their energies $E_n$}
\label{En}
\end{table*}

For each integer filling $0<n<8$ we choose the chemical potential $\mu$ from
the condition $E_{n-1}=E_{n+1}$ and consider the excitation energy $\Delta
E_{n}=E_{n+1}-E_{n}$ of adding or removing one electron. This way we obtain
$\Delta E_{2,4,6}=(U+J_{z})/2,$ while $\Delta E_{3,7}=(U^{\prime}-J_{z})/2.$

\subsection{Calculation of the non-local susceptibilities}

To study spin- and charge correlations we calculate the non-local charge (spin) susceptibility (cf. Refs. \cite{DB,MyEDMFT,AbInitioDGA,OurRev,MyEDMFT})
\begin{align}
\mathcal{\chi}_{q}^{c(s),MN,M'N'}&=\sum_{\nu,\nu'}\left[  (\chi_{q,\nu}^{0,MN,M'N'})_{MN,M'N'}^{-1} \delta_{\nu\nu^{\prime}}\right.\label{chi_BS}\\
&-\left.
\Phi_{q,\nu\nu^{\prime}}
^{c(s),MN,M'N'}+{V}_q^{c(s),mn,m'n'}\right]  _{\nu MN,
\nu^{\prime} M' N'}^{-1},\notag%
\end{align}
where $M=(m\alpha)$ etc. is the combination of spot- and valley indexes, ${V}_q^{c,mn,m'n'}=2(\widetilde{V}_{mm'}({\bf q})-v(\iu \omega_n))\delta_{mn}\delta_{m'n'}$, $V_q^{s}=0$, the bare susceptibility
\begin{equation}
\chi^{0,MN,M'N'}_{ q,\nu}=-T\sum\limits_{\mathbf{k}}G^{NM'}_{k}G^{N'M}_{k+q}\label{chi0}
\end{equation}
is considered as a matrix with respect to composite indexes $M,N$ and $M',N'$, $q=({\bf q},\iu \omega_n)$, $G_k^{MN}=[(i\nu_n+\mu-\Sigma(i\nu_n)) I-H]_{MN}^{-1}$ is the non-local Green's function, $H$ is the tight-binding Hamiltonian, and $I$ is the identity matrix. The vertices $\Phi$ are evaluated from the local vertex $\Gamma$ via the Bethe-Salpeter equation
\begin{align}
{\Gamma}_{\omega,\nu \nu'}^{c(s),MN,M'N'}&=\left[  (\Phi_{\omega,\nu \nu' }^{c(s),MN,M'N'})_{\nu MN,\nu' M'N'}^{-1} \right.\label{local_BS}\\
&-\left.
 \chi_{\omega,\nu}^{0,MN,M'N'} \delta_{\nu\nu^{\prime}}\right]  _{\nu MN,
\nu^{\prime} M' N'}^{-1},\notag%
\end{align}
where $\chi_{\omega,\nu}^{0,MN,M'N'}=-T G^{\rm loc}(\iu \nu) G^{\rm loc}(\iu \nu+\iu \omega) \delta_{MN'} \delta_{M'N}$. In the local vertices $\Gamma^{c(s),MN,M'N'}_{\omega,\nu\nu'}$ and $\Phi^{c(s),MN,M'N'}_{\omega,\nu\nu'}$ the composite indexes $M,M'$ correspond to the incoming particles with frequencies $\nu,\nu'+\omega$, while the indexes $N,N'$ refer to the outgoing particles with frequencies $\nu+\omega,\nu'$. Finally, the local vertices $\Gamma^{c(s),MN,M'N'}_{\omega,\nu\nu'}$ are constructed from the vertices, extracted from single-impurity problem,
\begin{align}
\Gamma^{c(s),MN,M'N'}_{\omega,\nu\nu'}=\Gamma^{\uparrow\uparrow,MN,M'N'}_{\omega,\nu\nu'}\pm \Gamma^{\uparrow\downarrow,MN,M'N'}_{\omega,\nu\nu'},
\label{Gamma}
\end{align}
where the first and second spin index refers to the first ($M,N$) and second ($M'N'$) pair of composite indexes, respectively. The details on the numerical solution of Bethe-Salpeter equations is discussed in Ref. \cite{MyEDMFT}. Note that due to the density-density form of the interaction, the second term in Eq. (\ref{Gamma}) is nonzero only for $M=N$,$M'=N'$. At the same time, the first term is non-vanishing also for $M=N'$, $N=M'$. The contributions of these latter combinations are important for the obtained incommensurate order; in Fig. \ref{FigChiqcs} we show the result for the charge susceptibility at $n=2$ with only diagonal vertices ($M=N$,$M'=N'$) included. The maximum of the non-uniform susceptibility shifts to commensurate positions in that case. 

\begin{figure}[h!]
	\center{\includegraphics[width=0.4\linewidth]{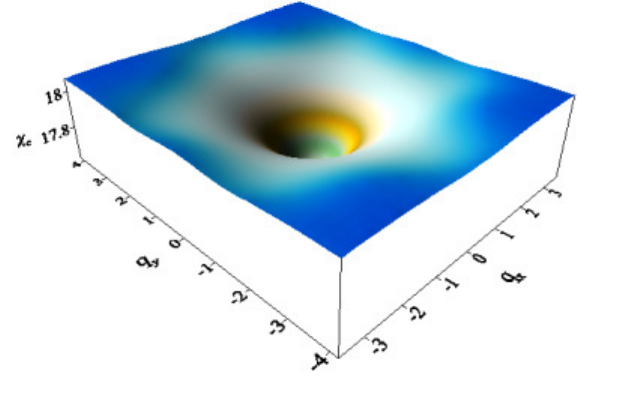}}
			\caption{(Color online) The momentum dependence of the staggered in valley indexes charge susceptibility for $n=2$ with only diagonal in pairs of composite indexes vertices included.}
	\label{FigChiqcs}
\end{figure}

\subsection{The momentum dependence of charge and spin susceptibility  }

Here we present additional results for the momentum dependence of charge and spin susceptibilities. In Fig. \ref{FigChiq1} we present the momentum dependence of the even in valley and spot indexes spin susceptibility $\chi_{\bf q}^{s}=\sum_{mn\alpha\beta} \mathcal{\chi}_{\bf q}^{s,m\alpha,n\beta}$ at $n=2$ (the staggered in valley spin susceptibility is almost equal to $\chi_{\bf q}^{s}$ due to almost vanishing inter valley components). One can see that its momentum dependence is quite similar to that for the staggered in valley charge susceptibility, see Fig. 4 of the paper, which supports emergent spin-valley symmetry scenario.

\begin{figure}[h!]
	\center{\includegraphics[width=0.45\linewidth]{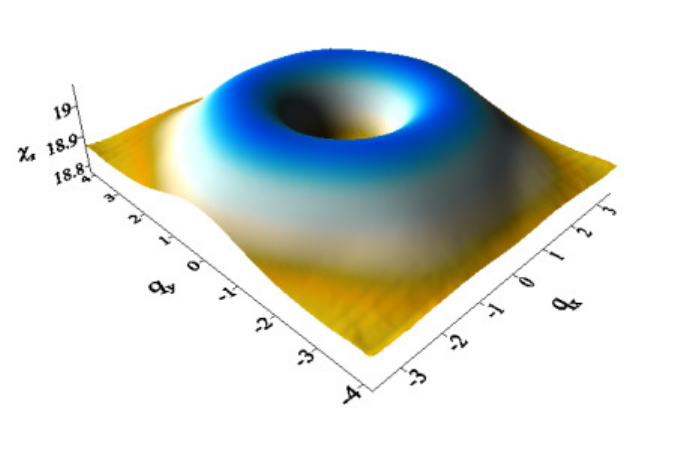}}
			\caption{(Color online) The momentum dependence of the even in spots and valleys spin susceptibility for $n=2$.}
	\label{FigChiq1}
\end{figure}

In Figs. \ref{FigChiq2} and \ref{FigChiq3} we present the momentum dependencies of charge and spin susceptibilities for $n=3$ and $n=4$. One can see that these dependencies are qualitatively the same, as for $n=2$, but on approaching half filling ($n=4$) the susceptibilities increase, while the wave vector of incommensurate correlations decreases, such that charge and spin correlations become more commensurate. 

Finally, in Figs. \ref{FigChiq5} and \ref{FigChiq6} we present the momentum dependencies of charge and spin susceptibilities for $n=5$ and $n=6$, which are similar to those for $n=3$ and $n=2$ due to approximate particle-hole symmetry. 

\begin{figure}[h!]
	\center{\includegraphics[width=0.85\linewidth]{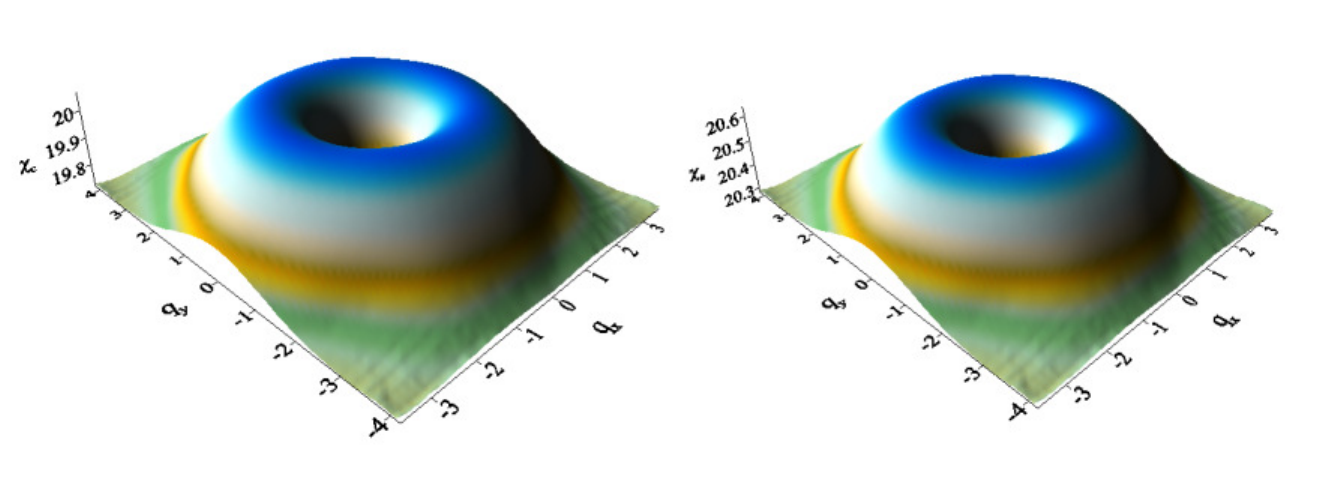}}
				\caption{(Color online) The momentum dependence of the staggered in valley charge susceptibility  (left) and even in spots and valleys spin susceptibility (right) for $n=3$.}
	\label{FigChiq2}
\end{figure}

\begin{figure}[h!]
	\center{\includegraphics[width=0.85\linewidth]{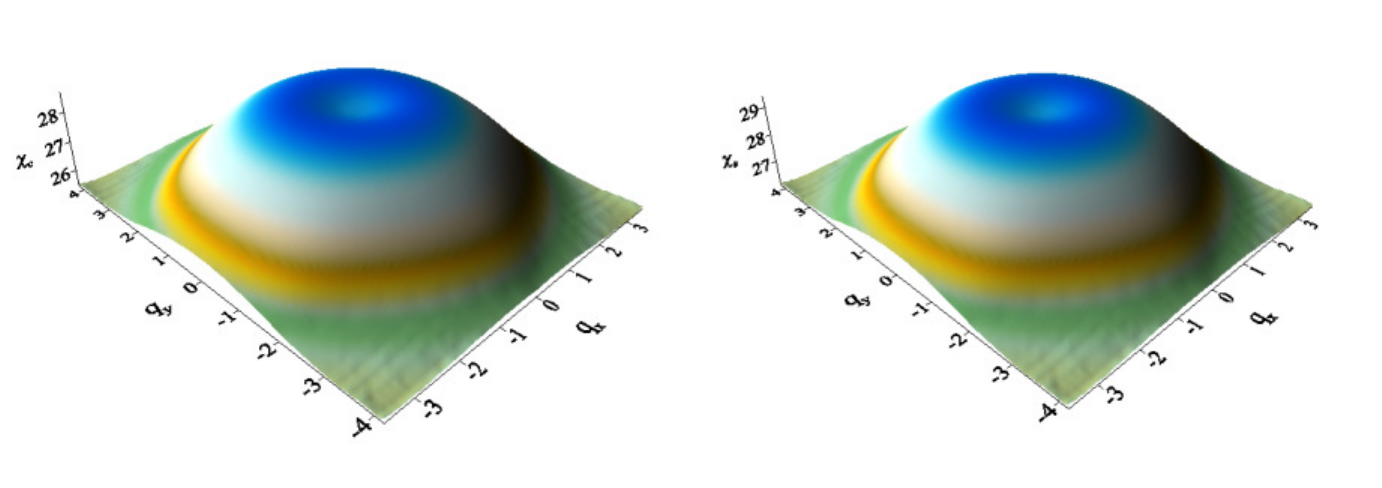}}
			\caption{(Color online) The momentum dependence of the staggered in valley charge susceptibility  (left) and even in spots and valleys spin susceptibility (right) for $n=4$.}
	\label{FigChiq3}
\end{figure}

\begin{figure}[h!]
	\center{\includegraphics[width=0.85\linewidth]{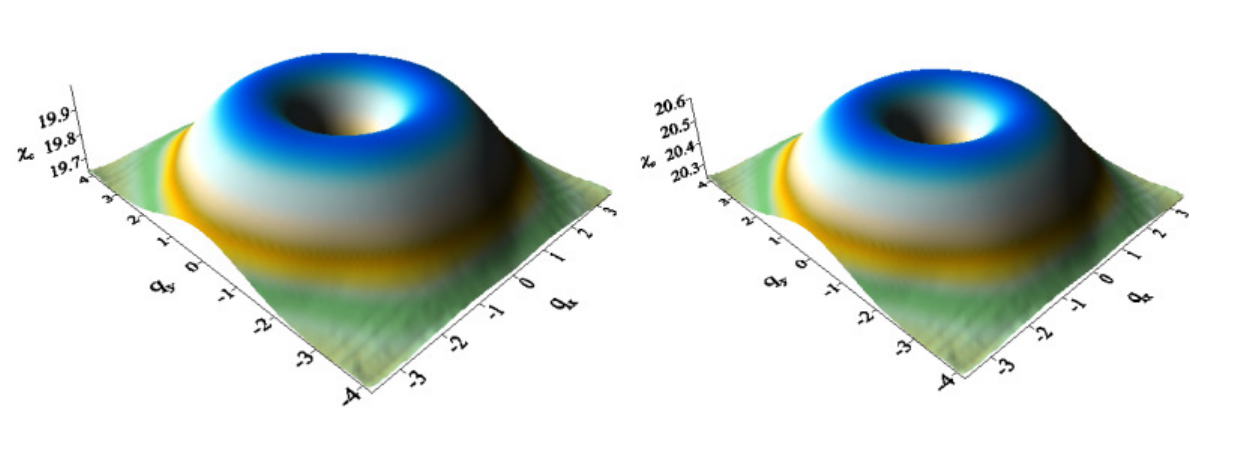}}
			\caption{(Color online) The momentum dependence of the staggered in valley charge susceptibility  (left) and even in spots and valleys spin susceptibility (right) for $n=5$.}
	\label{FigChiq5}
\end{figure}

\begin{figure}[h!]
	\center{\includegraphics[width=0.85\linewidth]{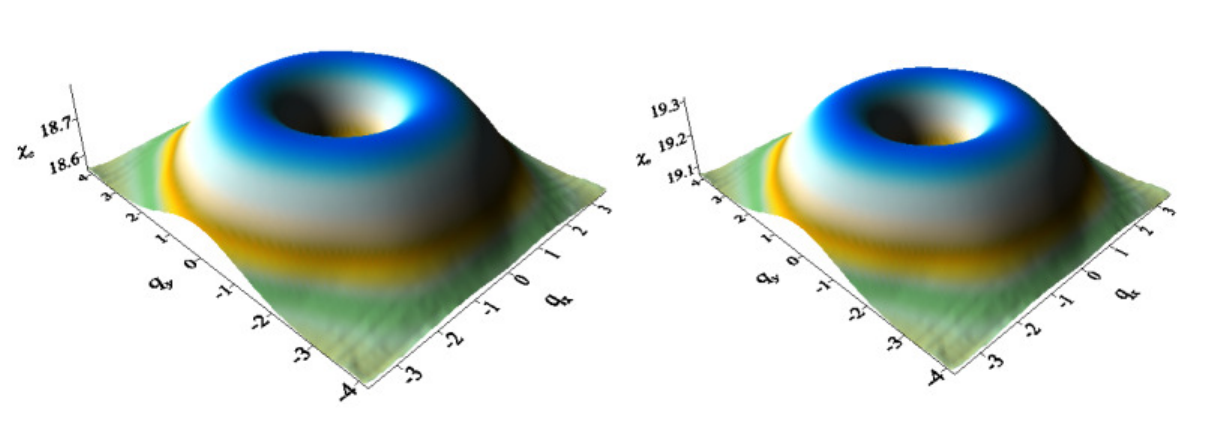}}
			\caption{(Color online) The momentum dependence of the staggered in valley charge susceptibility  (left) and even in spots and valleys spin susceptibility (right) for $n=6$.}
	\label{FigChiq6}
\end{figure}

\end{widetext}


\begin{thebibliography}{fe}
\bibitem{NoteJz} This value is somewhat smaller than obtained in Refs. \cite{ES10,SU41}, and accounts for the screening of Coulomb interaction in the low energy bands by the other bands.
\end{thebibliography}
\end{document}